# Investigation of Optical Spin Transfer Torque in Ferromagnetic Semiconductor GaMnAs by Magneto-Optical Pump-and-Probe Method

Eva Rozkotová[1], Petr Němec[1], Naďa Tesařová, František Trojánek[1], Petr Malý[1], Vít Novák[2], and Tomáš Jungwirth[2,3]

[1]Faculty of Mathematics and Physics, Charles University in Prague, Prague, 121 16, Czech Republic
[2]Institute of Physics ASCR v.v.i., Cukrovarnická 10, Prague, 162 53, Czech Republic
[3]School of Physics and Astronomy, University of Nottingham, Nottingham NG7 2RD, United Kingdom

We report on magnetization precession induced in (Ga,Mn)As by an optical spin transfer torque (OSTT). This phenomenon, which corresponds to a transfer of angular momentum from optically-injected spin-polarized electrons to magnetization, was predicted theoretically in 2004 and observed experimentally by our group in 2012. In this paper we provide experimental details about the observation of OSTT by a time-resolved pump-and-probe magneto-optical technique. In particular, we show that the precession of magnetization due to OSTT can be experimentally separated from that induced by the well known magnetic-anisotropy-related mechanism in a hybrid structure piezo-stressor/(Ga,Mn)As with an *in situ* electrical control of the magnetic anisotropy. We also illustrate that OSTT is clearly apparent in the measured dynamical magneto-optical signal in a large variety of (Ga,Mn)As samples with a Mn concentration from 3% to 9%.

## I. Introduction

Diluted magnetic semiconductors (DMSs), with (Ga,Mn)As as the most thoroughly investigated example, are materials that are prepared by a partial replacement of non-magnetic atoms by the magnetic ones [1], [2]. The Curie temperature achieved in these materials is still well below the room temperature [1], [2]. Nevertheless, their research is still rather appealing because it can provide fundamental insight into new physical phenomena that are present also in other types of magnetic materials where they can be exploited in realistic spintronic applications [1]-[3]. Optical spin transfer torque (OSTT) is a phenomenon where the spin angular momentum of carriers photo-injected in ferromagnetic semiconductor by absorption of circularly polarized light is transferred to the collective magnetization of the magnetic material. This effect was predicted theoretically several years ago [4], [5] and experimentally observed by our group very recently [6]. In this paper we provide further experimental details about this magneto-optical observation of OSTT [6]. In particular, we provide a detailed characterization of the hybrid structure piezo-stressor/(Ga,Mn)As where the *in situ* electrical control of the sample magnetic anisotropy enabled us to separate the magnetization precession induced by OSTT from that related to the pump-induced magnetic anisotropy change. We also show that OSTT can be clearly identified also in bare (Ga,Mn)As epilayers – we observed the corresponding signals in all investigated materials with a nominal Mn concentration from 3% to 9%.

## II. Experimental

We investigated the laser-pulse induced dynamics of magnetization by a pump-and-probe magneto-optical

Corresponding author: P. Němec (e-mail: nemec@karlov.mff.cuni.cz).

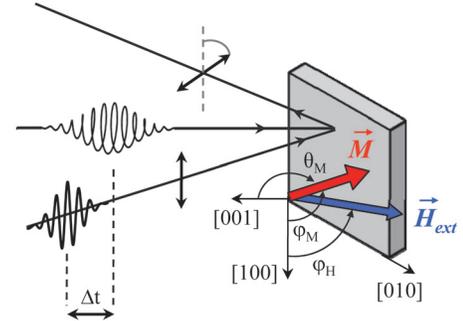

Fig. 1. Schematic diagram of the experimental set-up for a detection of the magnetization precession induced in (Ga,Mn)As by an impact of the circularly polarized femtosecond laser pump pulse. Rotation of the polarization plane of reflected linearly polarized probe pulses is measured as a function of the time delay $\Delta t$ between pump and probe pulses. The orientation of magnetization in the sample is described by the polar angle $\varphi_M$ and azimuthal angle $\theta_M$. The external magnetic field $H_{ext}$ is applied in the sample plane at an angle $\varphi_H$.

technique [7], [8]. A schematic diagram of the experimental set-up is shown in Fig. 1. The output of a femtosecond laser is divided into a strong pump pulse and a weak probe pulse that are focused to the same spot on the measured sample. Laser pulses, with the time width of 200 fs and the repetition rate of 82 MHz, were tuned to 1.64 eV, i.e. above the semiconductor band gap, in order to excite magnetization dynamics by photon absorption. The fluence of the pump pulses was from 46 to 70 μJ.cm$^{-2}$, which correspond to the photoinjected carrier density of $10^{18}$ cm$^{-3}$, and probe pulses were always twenty times weaker. The pump pulses were circularly polarized (with a helicity controlled by a wave plate) and the probe pulses were linearly polarized. The measured magneto-optical signals correspond to the probe polarization rotation induced by the pump pulses [8]. The experiment was performed close to the normal incidence geometry ($\theta_i = 2°$ and 8° for pump and probe pulses, respectively) with a sample placed in a cryostat, which was placed between the poles of an electromagnet. The external magnetic field $H_{ext}$ was applied in the sample plane at an angle $\varphi_H$ with respect to the [100] crystallographic direction in the sample plane (see Fig. 1). We investigated several (Ga,Mn)As samples from our set of high-quality epilayers, which are as close as possible to uniform uncompensated $Ga_{1-x}Mn_xAs$ mixed crystals [9], with a nominal Mn doping $x$ from 3% to 9% and Curie temperature $T_c$ from 77 to 179 K. The hybrid structures piezo-stressor/(Ga,Mn)As were prepared by gluing the (Ga,Mn)As sample with $x = 3.8\%$ to the commercial lead-zirconium-titanate piezo transducer (from Piezomechanik Gmbh) [10]. The strain generated in the sample by the piezo-stressor was measured in the two perpendicular directions by two-component strain gauge (from Vishay Micromeasurements Group) glued to the sample on the piezo-stressor.

### III. RESULTS AND DISCUSSION

It is a well known fact that an impact of a femtosecond laser pulse on (Ga,Mn)As can induce a precession of magnetization. However, in all the previous reports the measured dynamical magneto-optical (MO) signals were *independent on the pump pulse polarizations* [7], [8], [11]-[13]. Consequently, the underlying mechanism responsible for the magnetization precession initialization was identified as a sample magnetic anisotropy change by *a transfer of energy* from the pump pulse to the material [7], [11], [12]. In contrast, we have observed recently [6] that there is also a precession signal that is connected with *a transfer of angular momentum* from pump pulses. This is illustrated in Fig. 2 for a (Ga,Mn)As epilayer with a nominal doping $x = 3.8\%$ where the as-measured data for the circularly ($\sigma^+$ and $\sigma^-$) and linearly (s and p) polarized pump pulses are shown. We used piezo-stressor/(Ga,Mn)As hybrid structure that enabled us to control the sample magnetic anisotropy *in situ* (through the voltage dependence of a strain generated by the stressor [10]). By this we managed to eliminate the magnetic anisotropy-related mechanism of magnetization precession, which usually overshadows the signal due to OSTT. For the applied voltage $U = -150$ V, the precession of magnetization is observed only when circularly-polarized pump pulses are used - see Fig. 2(a). Moreover, a change of the circular polarization helicity results in a phase shift of 180° in the measured signal. On the other hand, for $U = +150$ V, where a sign of the strain is reversed, pump pulses with any polarization lead to the precession of magnetization but there is no obvious dependence of the precession phase on the polarization. The co-existence of two rather distinct excitation mechanisms, which both can lead to the precession of magnetization - but with a different initial phase, can be clearly revealed if the polarization-sensitive and polarization-insensitive parts of the signals are computed from the measured data – see Fig. 2(c)-(f). The polarization-insensitive part of the signal [Fig. 2(e) and (f)] is the same both for the linear and the circular polarization. This signal, which is connected with a magnetic anisotropy modification due to the pump-induced change of the hole concentration and of the sample temperature [7], [11], [12], is strongly dependent on the voltage applied to the piezo-stressor. In contrast, the polarization-sensitive signal shows sizable oscillations only when circular-polarization of pump pulses is used. Remarkably, this signal does not depend significantly on the applied voltage [see Fig. 2(c) and (d)]. We also point out that the observed oscilation periods for the polarization-independent and helicity-dependent excitations are identical. Independent whether the exitation was initiated by the shift of the magnetic easy axis or due to OSTT the precessing moments are the same in both cases.

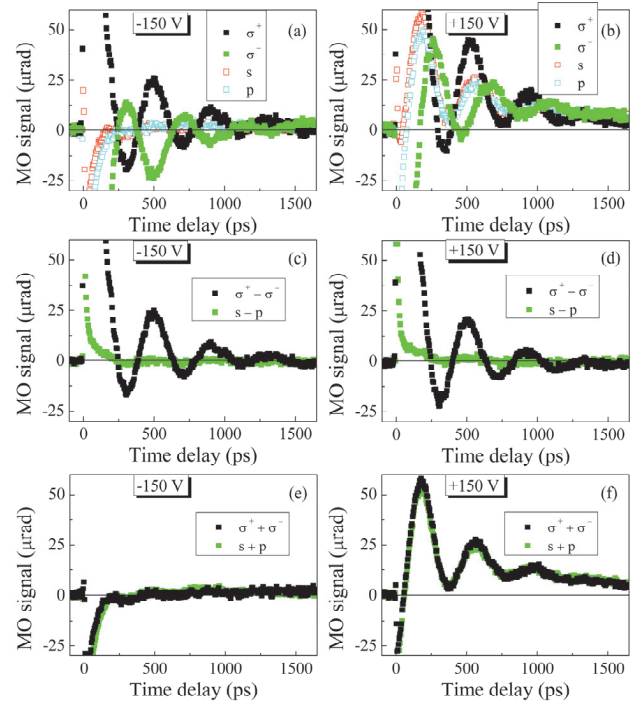

Fig. 2. Laser pulse-induced precession of magnetization in ferromagnetic (Ga,Mn)As epilayer with nominal doping $x = 3.8\%$ attached to a piezo-stressor. (a) and (b) Magneto-optical (MO) data measured for circularly ($\sigma^+$, $\sigma^-$) and linearly (s, p) polarized pump pulses for applied voltage $U = -150$ V and $U = +150$ V, respectively. Polarization-sensitive [$(\sigma^+ - \sigma^-)/2$ and $(s - p)/2$] and polarization-insensitive [$(\sigma^+ + \sigma^-)/2$ and $(s + p)/2$] parts of these signals are shown in (c), (d) and (e), (f), respectively. External magnetic field $\mu_0 H_{ext} = 30$ mT was applied along the piezo-stressor main axis ($\varphi_H = 115°$), sample temperature $T = 35$ K, pump intensity $I = 70$ μJ.cm$^{-2}$.

Due to the optical selection rules, absorption of circularly polarized pump pulse leads to a photoinjection of spin-polarized electrons and holes with a spin oriented along the direction of light propagation (i.e., perpendicular to the sample plane) [14]. Consequently, the carrier spins start to interact with ferromagnetic moments, which are oriented in the sample plane [9], via exchange coupling. This in turn leads to coupled precession dynamics of the magnetization and the photo-carrier spin density [4], [5], [6]. The considerably longer spin relaxation time of photo-electrons compared to that of photo-holes implies that the electrons have a much stronger influence on the magnetization orientation than the holes [4], [6]. We revealed by numerical modeling of the measured data by Landau-Lifshitz-Gilbert equation that the MO signals shown in Fig. 2 correspond to ≈ 1 degree out-of-plane tilt of magnetization from the equilibrium in-plane orientation [6].

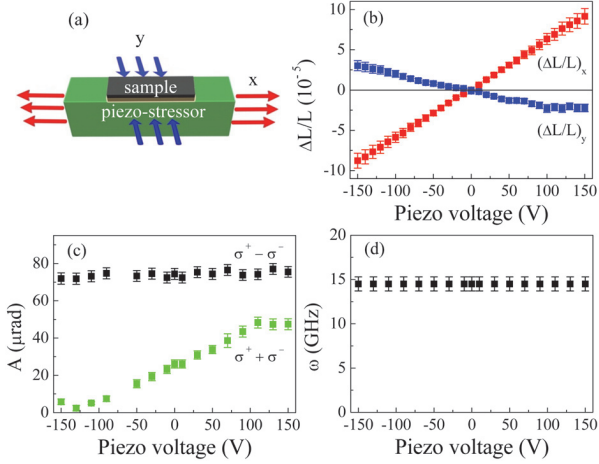

Fig. 3. Piezo-voltage control of laser pulse-induced precession of magnetization. (a) Schematic illustration of the piezo-stressor/(Ga,Mn)As hybrid structure at applied positive voltage; the piezo-stressor main axis was along the sample easy axis position (at the angle 115° from the [100] crystallographic direction in the sample). (b) Measurement of the strain generated by the piezo-stressor along its major ($x$) and minor ($y$) axes. Piezo-voltage dependence of the precession signal amplitudes $A$ (c) and angular frequencies $\omega$ (d) obtained by fitting the helicity-sensitive and helicity-insensitive parts of the measured MO signals; T = 35 K, $I$ = 70 μJ.cm$^{-2}$, $\mu_0 H_{ext}$ = 30 mT, $\varphi_H$ = 115°.

The influence of the voltage applied to the piezo-stressor is summarized in Fig. 3. The commercial piezo-stressors used in our experiments are formed by a large number of piezo ceramic/electrode layers that are electrically isolated from each other, and independently biased. The direction of the multilayer piling determines the main axis along which the device expands/compresses. The voltage-induced expansion in the major direction [$x$ in Fig. 3 (a)] is accompanied by a compression in the perpendicular directions – see Fig. 3 (b). The applied voltage $U = -150$ V induces a compressive strain of $9 \times 10^{-5}$ along the stressor main axis and a tensile strain of $3 \times 10^{-5}$ along the stressor minor axis. For $U = +150$ V the sign of the strains is reversed. We performed a measurement of the dynamical MO signals at various applied voltages. Each measured oscillatory MO signal was fitted by a damped harmonic signal superimposed on a pulse-like function [7], [8] from which we obtained the precession signal amplitude $A$ and angular precession frequency $\omega$. We observed that the application of the voltage to the stressor changes substantially the amplitude of the helicity-independent signal while it does not influence neither the amplitude of the helicity-dependent signal [see Fig. 3(c)] nor the precession frequency [see Fig. 3(d)]. The measured voltage dependence of the amplitude of the helicity-independent signal shows that the generated strains are sufficient for a modification of the sample magnetic anisotropy. In particular, the stressor (which is aligned with the easy-axis of the unstressed epilayer) makes the direction along its main axis magnetically easier for compressive strain, induced by negative $U$, and less easy for tensile strain, induced by positive $U$. Negative (positive) piezo-voltage, therefore, strengthens (weakens) the in-plane magnetic anisotropy easy-axis of the ferromagnetic semiconductor epilayer. This in turn leads to a reduced (enhanced) sensitivity of the easy axis position to the small pump-laser-induced change of the sample temperature and/or of the hole concentration [7] and, consequently, to a reduction (enhancement) of the precession amplitude for the helicity-independent signal [Fig. 3(c)]. On the other hand, the absolute magnitude of the sample magnetic anisotropy change is relatively small as apparent from the independence of $\omega$ on the piezo-voltage [see Fig. 3(d)].

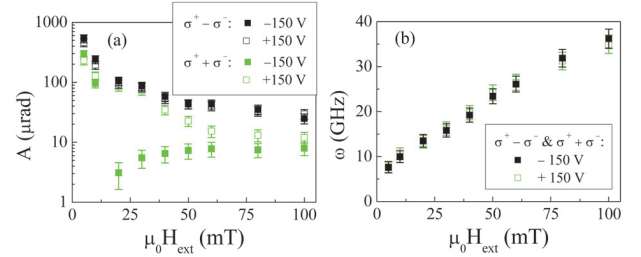

Fig. 4. External magnetic field control of precession amplitude $A$ (a) and angular frequency $\omega$ (b) for helicity-sensitive and helicity-insensitive parts of the measured MO signals at two piezo-voltages; T = 35 K, $I$ = 70 μJ.cm$^{-2}$, $\varphi_H$ = 115°. Note the logarithmic y-scale in (a).

In Fig. 4 we summarize the effect of the external magnetic field $H_{ext}$ applied along the stressor main axis on the precession signal. In Fig. 4(a) we show the dependence of $A$ on $H_{ext}$ for the helicity-sensitive and helicity-insensitive signals at positive and negative piezo-voltages. For helicity-sensitive oscillations at both voltages and for helicity-insensitive oscillations at $U = +150$ V, the amplitudes show the expected gradual decrease with increasing magnetic field [13]. On the other hand, for helicity-insensitive oscillations at $U = -150$ V the decrease is much more rapid. In fact, the helicity-insensitive oscillations are effectively quenched at 20 mT. As discussed above, this is because at large negative piezo-voltages the internal anisotropy field that is forcing magnetization to align with the stressor axis is significantly strengthened. With only a relatively weak additional external magnetic field, the changes in the magnetic anisotropy induced by the laser pulse are then not sufficient to trigger a sizable spin precession. The frequency of the magnetization precession is given by the magnetic anisotropy of (Ga,Mn)As as well as by the magnitude and direction of the external magnetic field [15]. The observed independence of $\omega$ on the piezo-voltage – see Fig. 4(b) – is in accord with Fig. 3(d) and with the fact that at higher magnetic fields the frequency is given mainly by the applied magnetic field and, therefore, the voltage applied to the piezo should not have any sizable effect.

In Fig. 5 we show the helicity-sensitive and helicity-insensitive signals measured in bare (Ga,Mn)As epilayers (i.e., in epilayers without the attached piezo-stressor). In fact, we were able to detect the helicity-sensitive signals in a vast majority of the investigated samples with different Mn concentration $x$. The relative strength of the helicity-sensitive signal with respect to that of the helicity-insensitive one depends strongly on the applied magnetic field and on the light intensity – the helicity-sensitive signal is typically from 5 to 20-times smaller.

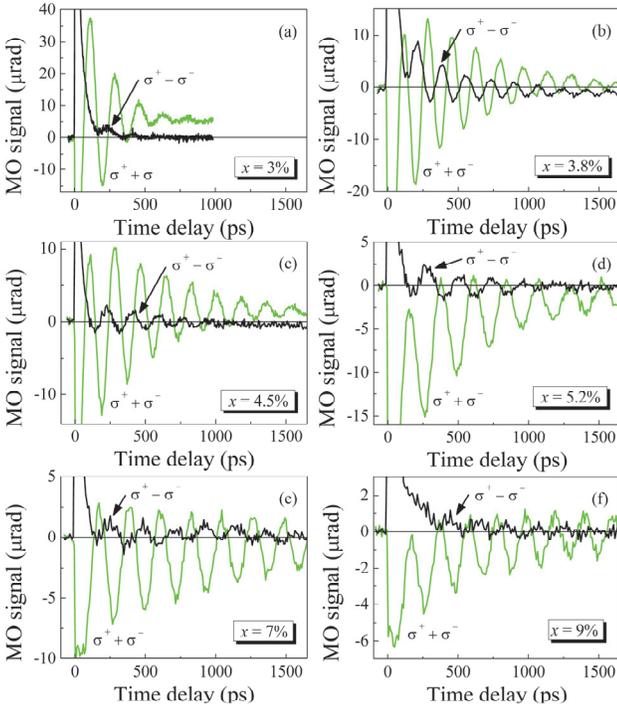

Fig. 5. Helicity-sensitive and helicity-insensitive parts of the MO signals measured in various $Ga_{1-x}Mn_xAs$ epilayers; $T = 15$ K, $I = 46$ μJ.cm$^{-2}$, $\mu_0 H_{ext} = 20$ mT, $\varphi_H = 90°$.

From the bare epilayers, the relative magnitude of the helicity-sensitive signal was the largest in sample with $x = 3.8\%$ where it was only about 2.5-times smaller than the helicity-insensitive one – see Fig. 5(b). Therefore, we selected this particular epilayer for the experiment with the piezo-stressor. Moreover, the attachment of the sample to the piezo itself led to a very strong enhancement of the helicity-sensitive signal which was 2.8-times *larger* than the helicity-insensitive one even for $U = 0$ V [see Fig. 3(c)]. This effect is a consequence from the fact that we glued the piezo-stressor to the sample at room temperature but the actual experiment was performed at low temperature. Therefore, due to the considerably smaller thermal contraction of the piezo-stressor compared to that of (Ga,Mn)As, the temperature decrease itself induced a rather strong tensile strain $\approx 10^{-3}$ even with no voltage applied to the piezo-stressor [10]. This additional strain reduced significantly the out-of-plane magnetic anisotropy of the (Ga,Mn)As that led to a considerable smaller precession frequency in the sample with the attached piezo-stressor (see Fig. 2) compared to that in the bare epilayer [see Fig. 5(b)]. Consequently, the reduced out-of-plane magnetic anisotropy enhanced the magnetization out-of-plane tilt due to OSTT.

## IV. Conclusion

Investigation of a transfer of angular momentum from spin polarized electrical current to the magnetization – spin transfer torque – opened a new research field of electrically driven magnetization dynamics which is currently aiming at a construction of a new generation of memory devices. The optical spin transfer torque, which we observed experimentally in ferromagnetic semiconductor (Ga,Mn)As, is a similar effect where the spin polarized carriers are generated optically. Nevertheless, one of the major differences between these two effect is connected with their rather distinct time scales – in electrical experiments the typical pulse lengths of several nanoseconds correspond to many magnetization precession periods while in the case of the optical excitation the torque is applied in the sub-picosecond timescale, which is much smaller that the precession period. Optical spin transfer torque can, therefore, open new research directions in the spin transfer physics where these phenomena will be investigated on the ultrafast time scale.

### Acknowledgment

This work was supported by the Grant Agency of the Czech Republic grant no. P204/12/0853 and 202/09/H041, by the Grant Agency of Charles University in Prague grant no. 443011 and SVV-2011-263306, by EU grant ERC Advanced Grant 268066 - 0MSPIN, and by Preamium Academiae of the Academy of Sciences of the Czech Republic.

### References

[1] T. Jungwirth, J. Sinova, J. Mašek, and A. H. MacDonald, "Theory of ferromagnetic (III,Mn)V semiconductors," *Rev. Mod. Phys.*, vol. 78, pp. 809-864, 2006.
[2] T. Dietl, "A ten-year perspective on dilute magnetic semiconductors and oxides," *Nature Materials*, vol. 9, pp. 965-974, 2010.
[3] Editorial, "More than just room temperature," *Nature Materials*, vol. 9, pp. 951, 2010.
[4] J. Fernández-Rossier, A. S. Núnez, M. Abolfath, and A. H. MacDonald, "Optical spin transfer in ferromagnetic semiconductors," http://arxiv.org/abs/cond-mat/0304492.
[5] A. S. Núnez, J. Fernández-Rossier, M. Abolfath, and A. H. MacDonald, "Optical control of the magnetization damping in ferromagnetic semiconductors," *J. Magn. Magn. Mater.*, vol. 272-276, pp. 1913-1914, 2004.
[6] P. Němec, E. Rozkotová, N. Tesařová, F. Trojánek, E. De Ranieri, K. Olejník, J. Zemen, V. Novák, M. Cukr, P. Malý, and T. Jungwirth, "Experimental observation of the optical spin transfer torque," *Nat. Phys.*, in press, http://arxiv.org/abs/1201.1436.
[7] E. Rozkotová *et al.*, "Light-induced magnetization precession in GaMnAs," *Appl. Phys. Lett.*, vol. 92, p. 122507, 2008.
[8] E. Rozkotová *et al.*, "Laser-induced precession of magnetization in GaMnAs," *IEEE Trans. Magn.*, vol. 44, pp. 2674-2677, 2008.
[9] T. Jungwirth *et al.*, "Systematic study of Mn-doping trends in optical properties of (Ga,Mn)As," *Phys. Rev. Lett.*, vol. 105, p. 227201, 2010 and its *Supplementary material*.
[10] A. W. Rushforth *et al.*, "Voltage control of magnetocrystalline anisotropy in ferromagnetic-semiconductor/piezoelectric hybrid structures," *Phys. Rev. B*, vol. 78, p. 085314, 2008.
[11] Y. Hashimoto, S. Kobayashi, and H. Munekata, "Photoinduced precession of magnetization in ferromagnetic (Ga,Mn)As," *Phys. Rev. Lett.*, vol. 100, p. 067202, 2008.
[12] J. Qi *et al.*, "Ultrafast laser-induced coherent spin dynamics in ferromagnetic $Ga_{1-x}Mn_xAs$/GaAs structures," *Phys. Rev. B*, vol. 79, p. 085304, 2009.
[13] H. Takechi, A. Oiwa, K. Nomura, T. Kondo, and H. Munekata, "Light-induced precession of ferromagnetically coupled Mn spins in ferromagnetic (Ga,Mn)As," *phys. stat. sol. (c)*, vol. 3, 4267-4270, 2006.
[14] M. I. Dyakonov and V. I.Perel in *Optical orientation,* edited by F.Meier and B.Zakharchenya, Vol. 8 of Modern problems in condensed matter sciences, North-Holland, Amsterdam, 1984, Chap. 2.
[15] E. Rozkotová, P. Němec, N. Tesařová, P. Malý, V. Novák, K. Olejník, M. Cukr, T. Jungwirth, " Coherent control of magnetization precession in ferromagnetic semiconductor (Ga,Mn)As," *Appl. Phys. Lett.*, vol. 93, p. 232505, 2008.